\documentclass[aps,prl,twocolumn,superscriptaddress,showpacs]{revtex4}
\usepackage{amssymb}
\usepackage{amsmath}
\usepackage{graphicx}

\begin{document}

\title{Path to finding the critical thickness for memory in thin
ferroelectric films}
\author{A.M. Bratkovsky}
\affiliation{Hewlett-Packard Laboratories, 1501 Page Mill Road, Palo Alto, California
94304}
\author{A.P. Levanyuk}
\affiliation{Hewlett-Packard Laboratories, 1501 Page Mill Road, Palo Alto, California
94304}
\affiliation{Dept. Fiz. Mat. Cond., Universidad Autonoma de Madrid, Madrid 28049, Spain}
\date{\today }

\begin{abstract}
The finite screening length by real metallic electrodes, albeit very small ($%
<1${\AA }), results in finite depolarizing field that tends to split the
film into domains. In very thin ferroelectric films the domain structure
reduces to sinusoidal distribution of polarization considered first in the
1980s. We discuss the phase transition between this structure and a single
domain state and show that it is first order, if it exists at all. The
alternative possibility is that the single domain state at zero bias voltage
would be metastable for all temperatures in most systems. This scenario
defines a path towards solution to a problem of finding parameters of a
system that can sustain the ferroelectric memory over a desired period of
time.
\end{abstract}

\maketitle

Stability of ferroelectricity in ultrathin films with electrodes is a topic
of intense current interest, as illustrated by numerous papers that appeared
recently in the leading journals (see, e.g. \cite{ghosez03, ghosez05, Sai,
Tagfirst, perkohl07}). In this Letter, we show that all those papers have
little or no relevance to the possibility of the ferroelectric (FE)\ memory
since they do not actually address the core issues. We shall formulate the
actual problem that has to be resolved in order to answer the question at
hand, and indicate possible paths to its solution. Our present formulation
builds on an old paper by Chensky and Tarasenko (ChT) \cite{ChT82}, which,
unfortunately, is rarely mentioned in recent literature. ChT discussed
stability of the FE films with respect to small fluctuations of the single
and multidomain states in an electroded film of a uniaxial ferroelectric
with `dead' layers between the electrodes and the material, Fig.1a (inset).
It was already shown that their system satisfactorily models also the FE
films without a physical dead layer but real metallic electrodes \cite%
{BLapl06}, i.e. it considers the same type of a system as have the recent
Refs.\cite{ghosez03, ghosez05, Sai, Tagfirst, perkohl07}, which customarily
disregard the multidomain region.

The results of ChT and the above authors for stability of homogeneous FE
state in the films would only be correct if the phase transitions between
the single and multidomain states were second order. In fact, the character
of the transition between the ferroelectric single domain state and the
multidomain ones was not studied by ChT, and it is one of the main questions
we address in this Letter. We find that in the low-$T$ part of the phase
diagram the transition is first order. Thus, the stability of the FE state
found by ChT means, in fact, its metastability at least in some temperature
region. With account of our old result \cite{BLPRL00}, we expect that the
single domain state is metastable for \emph{all} temperatures or, if one
studies the FE states at a given temperature but for different film
thicknesses, at all film thicknesses. This question has to be studied in
more detail, but in any case, the key theoretical problem to be resolved in
order to predict feasibility of the FE memories is not that of existence of
a solution corresponding to the ferroelectric phase or demonstrating its
stability with respect to small fluctuations. For the case of metastability
of a single domain state, the key problem is calculating the \emph{escape
rate }from this metastable state. We are not aware of theoretical efforts to
solve this problem. Another aspect of the memory problem is to find the
conditions of \emph{absolute} stability of the single domain state under
zero external bias voltage.

The single domain FE state is evidently stable in the case of the ideal
metallic electrodes (here and below we mean, of course, the ideally
homogeneous systems.) For a non-zero dead layer thickness ($d$) this is not 
evident anymore, although it is hardly surprising that there exists a
\textquotedblleft minimal thickness" $d_{m}$ such that for $d<d_{m}$ the FE
single domain phase is stable in the ChT sense, i.e. with respect to small
fluctuations, in all the temperature range. The minimal thickness $d_{m}$
depends on parameters on the FE material, as well as on the dielectric
constant of the dead layer (or, in the case of real metallic electrode, on
parameters of both the electrode and the ferroelectric film.) The stability
in the ChT sense of single domain FE state does \emph{not} mean that this
state is absolutely stable. According to our result in Ref.~\cite{BLPRL00},
at any finite thickness the homogeneously polarized state is metastable far
enough from the phase transition (or for thick enough films.) However, it is
stable not very far from the transition or for very thin films. This
conclusion makes this case attractive from the point of view of memory
applications. Unfortunately, the experimental systems of today do not fall
into this category \cite{BLapl06}. It does not seem impossible, at least in
principle, to find a suitable system, but we shall not dwell on this issue
here, assuming below that $d>d_{m}.$

Similar to ChT\cite{ChT82}, we consider a uniaxial ferroelectric film under
the external bias voltage $U$ (Fig.~1a, inset). The free energy of such a
film is given by\cite{ChT82}:
\begin{eqnarray}
\tilde{F} &=&F_{0}+\int_{FE}dV\Bigl[\frac{A}{2}P_{z}^{2}+\frac{B}{4}%
P_{z}^{4}+\frac{1}{2}D_{ij}\left( \nabla _{\bot i}P_{z}\right) \left( \nabla
_{\bot j}P_{z}\right)   \notag \\
&&+\frac{1}{2}\eta \left( \partial _{z}P_{z}\right) ^{2}+\frac{1}{2}\kappa
P_{zb}^{2}+\frac{A_{\bot }}{2}P_{\bot }^{2}+\frac{E^{2}}{8\pi }\Bigr]  \notag
\\
&&+\int_{DL}dV\frac{\epsilon _{e}E^{2}}{8\pi }+QU,  \label{eq:LGD}
\end{eqnarray}%
where $F_{0}$ is the free energy of the system at $\boldsymbol{P}=0,$ with $%
P_{z}$ the ferroelectric (switchable) component of polarization, $P_{zb}$
the nonferroelectric part of the polarization perpendicular to the electrodes%
\cite{Tagfirst}, $\boldsymbol{P}_{\bot }$ the in-plane polarization, $%
A=\gamma \left( T-T_{c}\right) ,$ and $B,D_{ij},\eta =\mathrm{const,}$ $%
\boldsymbol{\nabla }_{\bot }\mathrm{=(\partial }_{x},\partial _{y})$ is the
gradient in the plane of the film, $i,j=x,y$, and we assume summation over
repeating indices, $A_{\bot }>0$, $\boldsymbol{E}$ the electric field, $%
\epsilon _{e}$ the dielectric constant of the electrode/dead layer (marked
DL), $Q\approx -p$ the electrode charge, $p=\left\langle P_{z}\right\rangle $
is the average polarization in the FE film. We assume that $D_{ij}=D\delta
_{ij}$, which is valid, in particular, for BaTiO$_{3}$ or PbTiO$_{3}$ films
grown on (100) SrTiO$_{3}$ substrate because of a square symmetry in the
film plane. The noncritical in-plane and out-of-plane dielectric constants
are equal $\epsilon _{\perp }=1+4\pi P_{\perp }/E_{\perp }=1+4\pi /A_{\perp }
$ and $\epsilon _{b}=1+4\pi /\kappa $, respectively.

The stability of the paraelectric phase is lost with respect to appearance
of the \textquotedblleft polarization waves\textquotedblright\ \cite%
{ChT82,BLinh}:
\begin{equation}
\widetilde{P}_{z}(x,z)=a\cos qz\cos kx,  \label{eq:Pwave}
\end{equation}%
where, for $d\gtrsim 2d_{m}$ , $q\simeq \pi /l$ and $k=\left( 4\pi
^{3}/\epsilon _{\perp }Dl^{2}\right) ^{1/4}$\cite{ChT82,BLinh}. To find the
amplitude of the polarization wave in the external bias field, we use the
following approximation valid close to the phase transition \ (see below):
\begin{equation}
P_{z}(x,z)=p+\widetilde{P}=p+a\cos qz\cos kx.  \label{eq:pPtil}
\end{equation}%
In this case, the non-equilibrium free energy per unit area $\tilde{F}(p,a)$
takes the form:
\begin{equation}
l^{-1}\tilde{F}(p,a)=\frac{\tilde{A}+\xi }{2}p^{2}+\frac{\tilde{A}}{8}a^{2}+%
\frac{B}{4}p^{4}+\frac{3B}{8}a^{2}p^{2}+\frac{9B}{256}a^{4}-pE_{0},
\label{F}
\end{equation}%
where $\tilde{A}=A+2Dk^{2}=\gamma \left( T-T_{d}\right) ,$ $\xi =4\pi
d/\left( \epsilon _{e}l+\epsilon _{b}d\right) -2Dk^{2}\approx 4\pi d/\left(
\epsilon _{e}l\right) -2Dk^{2},$ $E_{0}=\epsilon _{e}U/\left( \epsilon
_{e}l+\epsilon _{b}d\right) \approx U/l$ the external field for the usual
case of a thin dead layer $\epsilon _{e}l\gg \epsilon _{b}d$. At $\xi >0$
(or $d>d_{m}$), according to the above expression for the free energy, the
system will undergo a phase transition at $E_{0}=0$ and $\tilde{A}=y=0,$
i.e. at $T=T_{d}$ into the sinusoidal domain phase. At lower temperatures,
there may be another transition into a homogeneously polarized state with $%
a=0,$ $p\neq 0$. The above potential allows one to study both phase
transitions also at non-zero external field.

When writing Eq.~(\ref{eq:pPtil}), we have assumed the stripe-like
sinusoidal domain structure. This is far from being obvious in our isotropic
case, and ChT \cite{ChT82} discussed possibilities of checkerboard and
hexagonal sinusoidal domain structures. However, those become irrelevant if
one takes into account the elastic strains. Their coupling to the
inhomogeneous polarization produces anharmonic terms in Eq.~(\ref{F}), with
the renormalized coefficients that now depend on the direction of the
\textquotedblleft wave vector" $k$ for any elastically anisotropic medium
\cite{Ema}. For the case of tetragonal uniaxial ferroelectric, the square
symmetry tells us that there are at least two orthogonal orientations of
stable stripe structures with the same free energy, while a hexagonal domain
pattern is clearly impossible. The checkerboard domain lattice should be
studied separately, but it is unlikely to be relevant. In the following, we
discuss a stripe structure neglecting the elasticity and assuming that its
role reduces mainly to selecting the direction of the sinusoidal
polarization waves. To be precise, elastic coupling also leads to changes of
the coefficient $B\ $in (\ref{F}), which are slightly different between the
third, fourth, and fifth terms there (see Ref.\cite{Ema}), but this is only
a numerical difference that does not affect any of the results below, and we
shall not dwell on this issue here.

By minimizing the free energy (\ref{F}) with respect to $a,$ one finds the
equilibrium amplitude of the wave:
\begin{equation}
a_{0}=\left[ -\frac{16}{9B}\left( \tilde{A}+3Bp^{2}\right) \right] ^{1/2}=%
\frac{4p_{c}}{3^{1/2}}\sqrt{1-s^{2}},  \label{P0eq}
\end{equation}%
where $s=p/p_{c},$ with $p_{c}=\sqrt{-\tilde{A}/3B}$ the characteristic
polarization. We finally arrive at the dimensionless free energy, $f=3B\xi
l^{-1}\tilde{F}$:

\begin{equation}
f_{\pm }=\left\{
\begin{array}{cc}
\frac{1}{2}y\left( 1-y\right) s^{2}+\frac{1}{12}y^{2}s^{4}-\sqrt{y}se, &
\left\vert s\right\vert \geq 1, \\
-\frac{y^{2}}{3}+\frac{y}{2}\left( 1+\frac{y}{3}\right) s^{2}-\frac{1}{4}%
y^{2}s^{4}-\sqrt{y}se, & \left\vert s\right\vert <1,%
\end{array}%
\right.  \label{eq:fpm}
\end{equation}%
where $e=E_{0}/\zeta $ the relative external field, $\zeta =\xi ^{3/2}/\sqrt{%
3B}$ the characteristic electric field, and
\begin{equation}
y=-\widetilde{A}/\xi ,  \label{eq:y}
\end{equation}%
the characteristic temperature (i.e. the relative distance of transition
temperature from the paraelectric phase that depends on the film thickness $%
l $). It is easy to see that the free energy is continuous with the first
derivative with respect to $s,$ while $d^{2}f_{-}/ds^{2}\neq
d^{2}f_{+}/ds^{2}$ at $s=\pm 1$.

One of the easiest ways to reveal the order of the transition is to inspect
the equations of state curves $s=s\left( e\right) $ obtained \ from the
condition $df/ds=0,$ which read:
\begin{equation}
\sqrt{y}\left( 1-y\right) s+\frac{1}{3}y^{3/2}s^{3}=e,\qquad \left\vert
s\right\vert \geq 1  \label{eq:xplus}
\end{equation}%
\begin{equation}
\sqrt{y}\left( 1+\frac{y}{3}\right) s-y^{3/2}s^{3}=e,\qquad \left\vert
s\right\vert <1.  \label{eq:xminus}
\end{equation}%
Recall that the given state is (meta)stable only when $d^{2}f/ds^{2}>0.$
Several typical $s\left( e\right) $ curves for $y=1/4,$ $1,$ and $3$ are
displayed in Fig.~1.
\begin{figure}[h]
\begin{centering}
\includegraphics
[width=5cm]{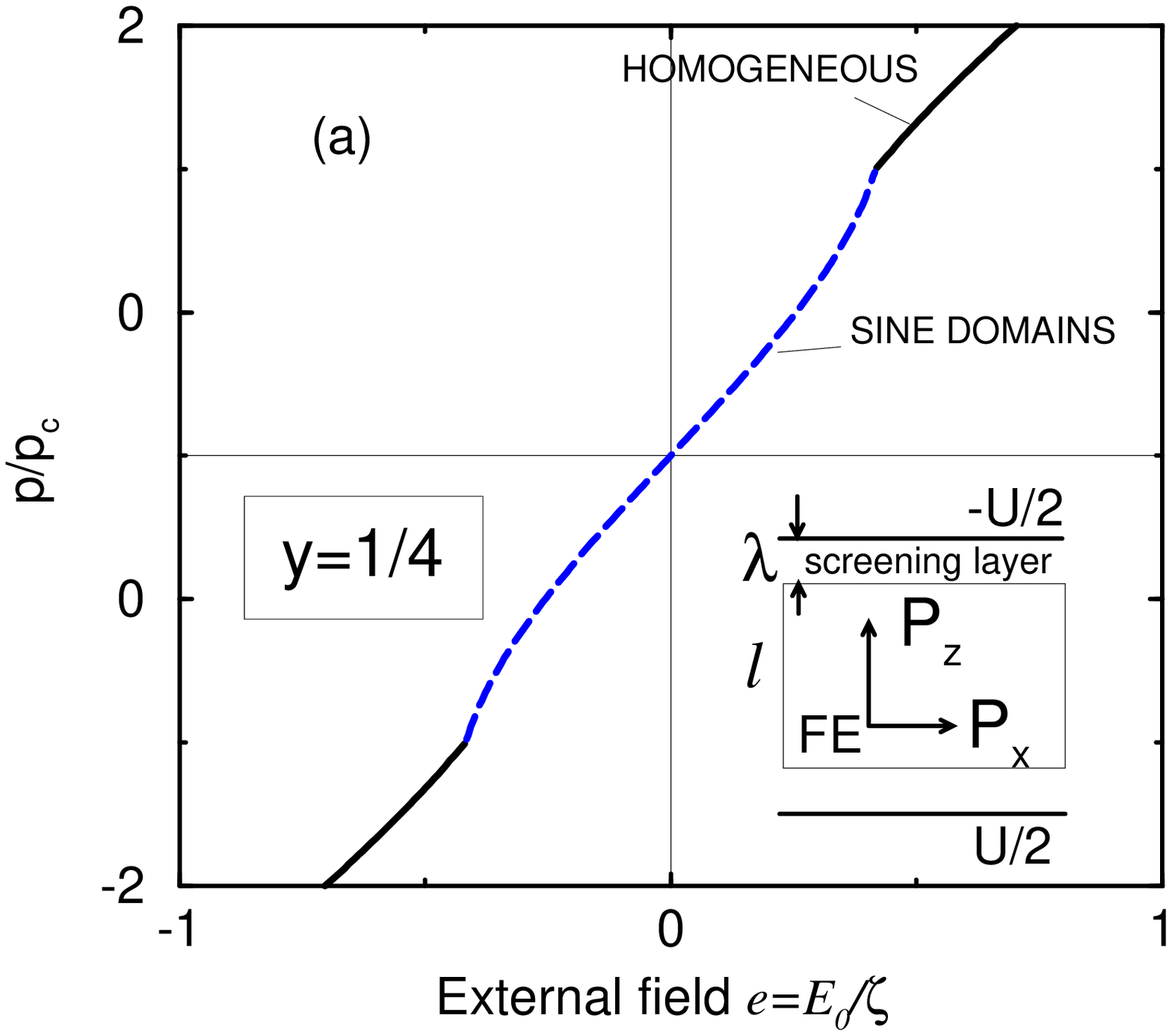}
\includegraphics
[width=5cm]{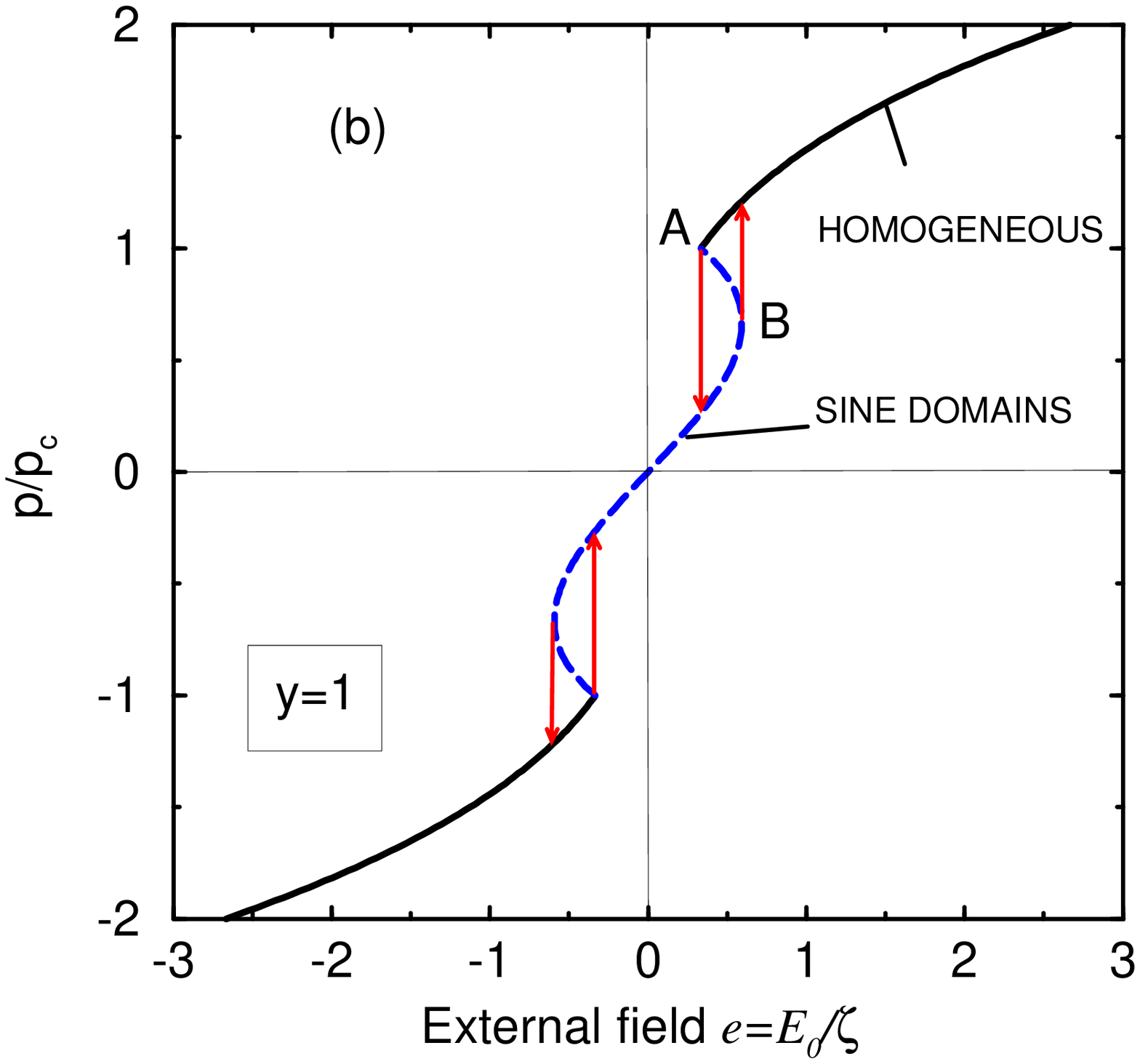}
\includegraphics
[width=5cm]{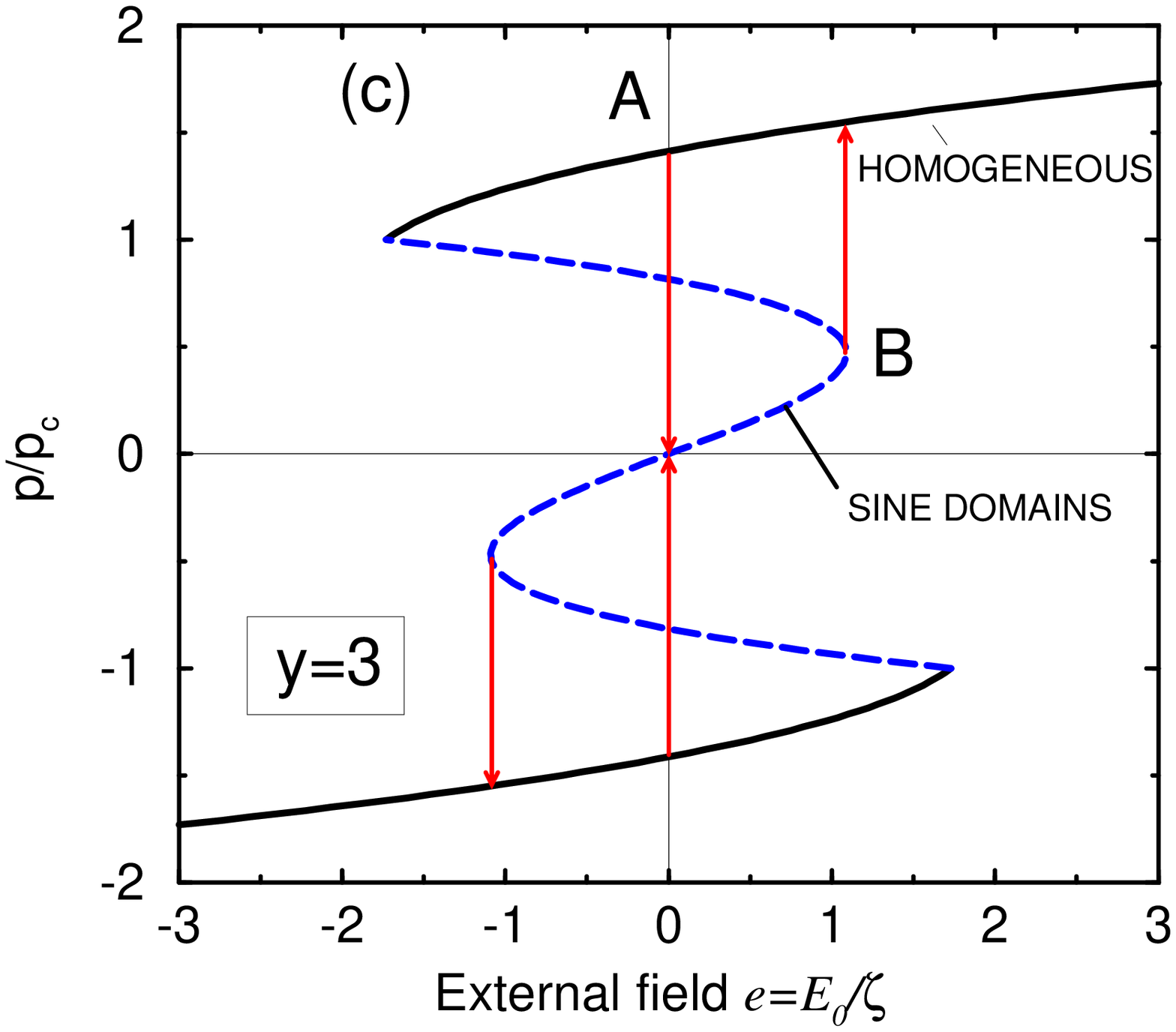}
\caption{The equation of state
$p=p(E_0)$ for ferroelectric film with dead layer in external field
$E_0$ for various values of the relative temperature $y$,
Eq.~(\ref{eq:y}): (a) $y=1/4$, (b) $y=1$, and (c) $y=3$. There is
a phase transition between the homogeneous state and the one with
sinusoidal domains: (a) second order, which becomes first order in
cases (b) and (c), where arrows indicate the hysteretic behavior during
polarization switching. The terminal external fields for hysteresis are
at points A and B.  All those cases are very different from the FE film
without the dead layer. $\zeta=\xi ^{3/2}/\sqrt{3B}$
is the characteristic electric field.
Inset (a) shows the schematic of the ferroelectric film with either real electrodes or
the dead layer with the thickness $\lambda$.
}\label{fig:eqs}
\end{centering}
\vspace{0.05 cm}
\end{figure}
We see that at a relatively small $y=1/4$ (not far below $T_{d}$ or $\tilde{A%
}=y=0$) and in the field $e>e_{c}=\sqrt{y}-2y^{3/2}/3=5/12$ the system is
homogeneously polarized. In the lower bias field, it splits via the second
order phase transition into domains with zero net polarization at $E_{0}=0.$
At $y>3/8,$ the transition is first order, which is clearly seen for $y=1,$
where two metastable solutions exist in the fields $e_{A}<E_{0}/\zeta <e_{B}$%
, Fig.~1. The points of the first order phase transitions found from the
condition of equal free energies, $f_{+}[s_{+}(e)]=$ $f_{-}[s_{-}(e)],$ are
marked in Fig.~1. The first instance when the state with spontaneous net
polarization $p\neq 0$ at $E_{0}=0$ becomes \emph{formally} possible as a
solution to the equations of state is at $y=3/2$. However, this state is
unstable ($d^{2}f_{-}/ds^{2}<0$ at $s=\pm 1),$ as is evident from the
negative slope of the lower branch of $p-e$ curve at $s=1$ in Fig.~1b in the
external fields $e_{A}<E_{0}/\zeta <e_{B}.$
\begin{figure}[h]
\begin{centering}
\includegraphics
[width=5cm]{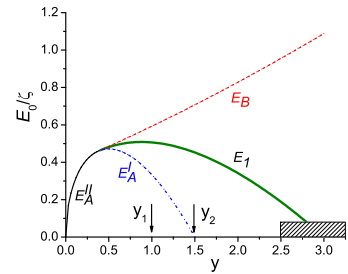}
\caption{
External field-temperature (thickness) phase diagram of a ferroelectric thin
film with real metallic electrodes.
The line of stability loss of the homogeneous states ($E_{A}^{I},E_{A}^{II}$)
has been calculated in Ref.~\cite{ChT82}. The tricritical point ($y=3/8$) and
lines $E_{1}$  (line of first order phase transitions), 
$E_{B\text{ }}$ are from the present work. The point $y_1=1$
corresponds to the "critical thickness" calculated in Refs.~\cite{ghosez03,
ghosez05, Sai, Tagfirst} with an account for atomistic structure of
the electrodes (additional boundary conditions), $y_2=3/2$ is the stability
point considered in Ref.~\cite{perkohl07} with an account for
elastic strains. The shaded region is area of interest for estimating the
retention time  of homogeneously polarized state.
}\label{fig:Chdiag}
\end{centering}
\vspace{0.05 cm}
\end{figure}
The above tricritical behavior of thin ferroelectric films with real
electrodes becomes clear from the phase diagram in $(E_0,y)$ plane, Fig.~2.
Indeed, there is a second order phase transition in the interval $0<y<3/8,$
where the sinusoidal domains form in the fields below $e=e_{A}$, where
\begin{equation}
e_{A}=E_{A}/\zeta =\sqrt{y}(1-2y/3),  \label{eq:EA}
\end{equation}%
corresponding to point A in Fig.~1b. In the range $3/8<y<3/2,$ the same
expression gives the line of stability loss of the single domain phase.
After passing the tricritical point at $y=3/8,$ there appears a line of
first order phase transitions $E_{1}(y),$ that terminates at $y=3,$ and it
is bracketed by the lower and higher terminal fields $e_{A}$ (\ref{eq:EA})\
and $e_{B}$, where%
\begin{equation}
e_{B}=E_{B}/\zeta =2(3+y)^{3/2}/27.  \label{eq:EB}
\end{equation}%
The physical meaning of these fields is clear from Fig.~2:\ $E_{A}$ is the
lowest field where the stability of the homogeneous phase is lost with
respect to domains, while $E_{B}$ is the highest field where the domain
structure still exists. The hysteresis, therefore, is observed in the range
of fields $E_{A}<E_{0}<E_{B},$ when $y>3/8$. Although the approximation of
sinusoidal domains becomes rather poor at larger parameters $y\gtrsim
y_{1}=1,$ it should still correctly grasp the main features of the phase
behavior. Note that this result invalidates a long unchallenged claim by
Chensky and Tarasenko\cite{ChT82} that one can prepare a monodomain state at $E_{0}=0,$ $%
\ y>y_{2}=3/2$ (that corresponds to low temperatures well below $T_{c}$ and $%
T_{d}$) by polarizing the system in sufficiently high field and then
removing the field.

The present diagram suggests that the homogeneously polarized state will
remain metastable (in the present one-sinusoid approximation)\ only at $y>3$%
. There is a metastability of homogeneously polarized state in the region $%
3/2<y<3$. Formally, both the ferroelectric memory and the polarization
switching are possible at these temperatures/thicknesses but no conclusion
of practical importance can be made before calculating the \emph{escape time}
from the metastable state. At larger $y\gtrsim 3$ (further down from the
phase transition with respect to temperature, or for films with thickness
exceeding the critical one)\ the state with the homogeneous polarization in
the present single-harmonic approximation, Eq.~(\ref{eq:pPtil}), has the
same or \emph{lower} free energy than the state with $p=0:$ $%
f_{+}(s_{m})\leq f_{-}(0),$ where $\pm s_{m}$ are the positions of the
minima of the free energy $f_{+}$ at $E_{0}=0$ (Fig.~3). Note that this
result is approximate. The reason is that Eq.~(\ref{eq:pPtil}) is valid near
the phase transition point only. The region of validity of this
approximation has been estimated in \cite{ChT82} as roughly $-\tilde{A}%
<Dk_{c}^{2}$, which means $y\lesssim 1$ if $4\pi d/\left( \epsilon
_{e}l\right) $ and $2Dk_{c}^{2}$ are of the same order of magnitude.

We should mention that the second order phase transition into \emph{%
homogeneous} FE phase, considered in the prior papers \cite{ghosez03,
ghosez05, Sai, Tagfirst}, may only occur in zero field, $E_{0}=0,$ and it
corresponds to $A=4\pi d/\left( \epsilon _{e}l+\epsilon _{b}d\right) ,$ or,
in other words, to the point $y=y_{1}=1$ on the $(E_{0},y)\ $diagram, which
is the point that in typical circumstances lies well inside the domain
regime and, therefore, is normally unreachable, Fig.~2. On the other hand,
Pertsev and Kohlstedt\cite{perkohl07} have studied the stability loss of the
FE homogenous state, which corresponds to point $y=y_{2}=3/2$ in the phase
diagram. They noted that one should take elastic coupling into account while
discussing the stability of the ferroelectric state but, unfortunately, did
it in a confusing manner with incorrect conclusions, see analysis in \cite%
{BLcomPer07}.
\begin{figure}[h]
\begin{centering}
\includegraphics
[width=5cm]{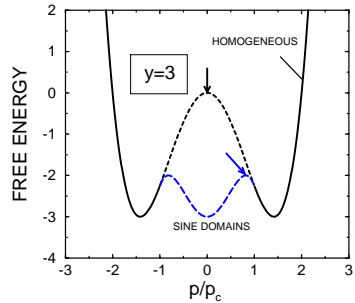} \caption{The free energy of the FE
film with the dead layer for the relative temperature $y=3$, when
the transition between the homogeneously polarized state and
sinusoidal domains is first order. It is evident that switching
proceeds through the state with domains with much lower energy
barrier for nucleation (top of the barriers indicated by arrows). An
account for higher harmonics in Eq.~(\ref{eq:Pwave}) will deepen the
energy of the domain state compared with the one shown here and will
lower the barrier even further.}\label{fig:Free}
\end{centering}
\vspace{0.05 cm}
\end{figure}
In a more accurate approximation accounting for higher harmonics to describe
the inhomogeneous polarization, the free energy minimum at $s=E_{0}=0$ dips
lower than that of the homogeneous state. It is those higher harmonics that
convert the sinusoidal domain structure into a conventional one with narrow
domain walls. For the parts of the curves corresponding to $\left\vert
s\right\vert \gtrsim 1,$ these higher harmonics are not important (they are
when an amplitude of the first harmonic becomes substantial) but they will
change the curves for $\left\vert s\right\vert <1$ substantially. The
amplitudes of the higher harmonics are to be considered as new variational
parameters for the free energy, and their account will be lowering the
estimated free energy. Hence, the minimum at $s=p=0$ in Fig.~3 is actually
deeper, and the homogeneously polarized phase becomes stable not at $y=3$
but at a larger value (i.e. at a lower temperature or a \emph{larger
thickness}.) Furthermore, it is possible that the homogeneous state would
always remain less stable than the polydomain state in that region. Indeed,
in the opposite limiting case, i.e. far below the FE transition, it has been
shown that for any thickness of the dead layer the multidomain state has
lower free energy than the homogeneously polarized state \cite{BLPRL00}.
Anyway, what value of $y$ would correspond to the \textquotedblleft critical
thickness for the ferroelectricity" within the discussed simple case depends
on the desirable memory retention time and should be found by solving a
kinetic problem.

The present discussion of an equilibrium problem in the one-sinusoid
approximation can be improved. We have already mentioned the effect of
coupling of the polarization and the elastic strains, which is unimportant
for studying the stability of the paraelectric phase but should be taken
into account while considering the multidomain state. This is valid also for
the discussion of stability of a single domain FE phase. We can mention also
the effect of the additional boundary conditions, apart from electrostatic
ones used by ChT, and the effects of higher order terms in the LGD
expansion. This, however, would not change the present qualitative
conclusions that will apply also to finite FE patches with lateral sizes of $%
100$nm and smaller, which have lateral dimensions still much larger than the
domain width, which is just $\sim 1-2$nm in $5$nm thick BaTiO3 film\cite%
{BLapl06}. Our discussion above indicates only the first steps on the path
to addressing the kinetic problem in a likely scenario that the single
domain state is not absolutely stable but metastable.

APL has been partially supported by Spain's MEC under Grant
NAN2004-09183-C10-05.

\end{document}